\documentclass[preprint,nofootinbib]{revtex4}
\usepackage{graphicx}
\usepackage{dcolumn}
\usepackage{amsmath,amsthm,amssymb}
\usepackage{epstopdf}
\usepackage{mathrsfs}

\DeclareSymbolFontAlphabet{\mathrsfs}{rsfs}
\DeclareMathAlphabet{\mathcal}{OMS}{cmsy}{m}{n}

\newcommand{\be}{\begin{equation}}
\newcommand{\ee}{\end{equation}}

\begin{document}

\title{Saddle-point dynamics of a Yang-Mills field on the exterior Schwarzschild spacetime}

\author{Piotr Bizo\'n}
\affiliation{M. Smoluchowski Institute of Physics, Jagiellonian University, Krak\'ow, Poland}
\author{Andrzej Rostworowski}
\affiliation{M. Smoluchowski Institute of Physics, Jagiellonian University, Krak\'ow, Poland}
\author{An\i l Zengino\u{g}lu}
\affiliation{M. Smoluchowski Institute of Physics, Jagiellonian University, Krak\'ow, Poland}
\date{\today}
\begin{abstract}
We consider the Cauchy problem for a spherically symmetric $SU(2)$ Yang-Mills field propagating outside the Schwarzschild black hole. Although solutions starting from smooth finite energy initial data remain smooth for all times, not all of them scatter since there are non-generic solutions which asymptotically tend to unstable static solutions.
 We show that a static solution with one unstable mode appears as an intermediate attractor in the evolution of initial data near a border between basins of attraction of two different vacuum states.
 We study the saddle-point dynamics near this attractor, in particular we identify the universal phases of evolution: the ringdown approach, the exponential departure, and the eventual decay to one of the vacuum states.
\end{abstract}
\maketitle
\section{Introduction}
It is well known that solutions of  Yang-Mills  equations in four dimensional Minkowski spacetime are globally regular in time for reasonable initial data. This fact was first proved by Eardley and Moncrief for smooth  data \cite{em} and later strengthened by Klainerman and Machedon for finite energy  data \cite{km}.
Due to the work of Christodoulou \cite{ch} and Glassey and Strauss \cite{gs} it is also known that the energy of the Yang-Mills field in any bounded region of flat spacetime decays to zero and that all solutions scatter.

A natural question is whether these properties remain valid in a curved background. Part of this question was answered affirmatively by Chru\'sciel and Shatah who proved global regularity of Yang-Mills equations on arbitrary globally hyperbolic four dimensional spacetimes \cite{cs}. However, studies of Yang-Mills equations on the Schwarzschild background revealed a key difference in comparison with the flat spacetime case: the equations admit static solutions \cite{bc,b1}, which implies that not all solutions scatter.

The quantitative behavior of Yang-Mills fields on curved spacetimes seems under-explored. The only results we are aware of are concerned with late-time tails for small data solutions on the Schwarzschild background \cite{bcr,z1}.

The aim of this paper is to examine the Cauchy problem for large data solutions of the Yang-Mills equations on the exterior Schwarzschild background.  We investigate thoroughly the role static solutions play in the dynamics. These solutions are unstable and therefore they are not observed in the Cauchy development of generic initial data. We show, however, that static solutions do participate in the evolution of specially prepared initial data. In particular, the static solution with one unstable mode appears as an intermediate attractor in the evolution of initial data  near a border between basins of attraction of two different copies of the vacuum solution. Solutions with data lying on this borderline tend asymptotically to the static solution and therefore they do not scatter.

This behavior bears many similarities with critical phenomena in gravitational collapse for the coupled Einstein-Yang-Mills system \cite{ccb,chm}, where the colored black hole with one unstable mode appears as an intermediate attractor. In the case of fixed background the problem is much simpler and thus amenable to a more detailed analytical description. In particular, we show that for intermediate times the convergence to the  static solution along the codimension-one stable manifold proceeds via quasinormal ringing. We look also at the nonlinear bi-instability of the perturbed attractor and compute the energy fluxes  through the horizon and null infinity during its decay to vacuum.

The rest of the paper is organized as follows. In Section~\ref{sec:prelimin} we introduce the model
and discuss its basic properties. In Section~\ref{sec:static} we discuss static solutions, their linearized instability and quasinormal modes. Our main results, based on mixed analytical and numerical arguments, are presented in Section~\ref{sec:numeric}. We conclude with general remarks in Section~\ref{sec:discuss}.
\section{Preliminaries}
\label{sec:prelimin}
\subsection{The Schwarzschild metric}
The Schwarzschild metric with mass $m>0$ in standard coordinates $(\bar{t},r, \vartheta, \varphi)$ reads
\be g=-N\,d\bar{t}^2 + N^{-1}\,dr^2+r^2\,(d\vartheta^2+\sin^2{\vartheta} d\varphi^2)\,, \quad \textrm{with}\quad N=1-\frac{2m}{r}. \ee
 We restrict our attention to the exterior region $r\geq r_h=2m$. Hypersurfaces of constant time coordinate all meet at the bifurcation sphere near the black hole and at spatial infinity in the asymptotic domain. This pathological behavior, which is inconvenient for certain applications, can be removed by introducing
 a new time coordinate
 \be\label{time_traf} t = \bar{t}-h(r)\,, \ee
 with a suitable height function $h$ (\cite{z3}). Note that the new foliation by $t$ respects the
stationarity of Schwarzschild spacetime, that is, the representation of the timelike Killing vector field of Schwarzschild spacetime is invariant under the transformation \eqref{time_traf}, i.e. $\partial_{\bar{t}}=\partial_t$.
The transformed metric takes the form
\be \label{eq:decomp} g = -N\,dt^2 - 2 \,NH\, dt\,dr +\frac{1-(NH)^2}{N}\,dr^2+r^2\,(d\vartheta^2+\sin^2{\vartheta} d\varphi^2)\,, \ee
where $H(r)=h'(r)$. Most commonly employed coordinatizations of the Schwarzschild spacetime lead to a metric of the above form. For example, the ingoing Eddington-Finkelstein coordinates correspond to $H = -\frac{1}{N} \frac{2m}{r}$.
This choice of the height function removes the pathological behavior at the horizon: the constant time hypersurfaces, instead of intersecting at the bifurcation sphere as for Schwarzschild coordinates,  foliate the event horizon. Consequently, all metric components are regular at the event horizon. The Painlev\'e-Gullstrand coordinates are similar with $H = -\frac{1}{N} \sqrt{\frac{2m}{r}}$.

Analogously, hyperboloidal foliations with a suitable asymptotic behavior avoid the pathological behavior at spatial infinity: the constant time hypersurfaces, instead of meeting at spatial infinity, foliate future null infinity \cite{h}. A special class of hyperboloidal foliations is given by constant mean curvature (CMC) hypersurfaces \cite{cmc}. For a suitable choice of parameters they lead to constant time hypersurfaces which are regular both at the event horizon and at infinity. The derivative of the height function for the CMC hypersurfaces reads \be\label{cmc}
H= \frac{1}{N} \frac{J} {\sqrt{J^2+N}}, \quad \textrm{with} \quad J
= \frac{K r}{3}-\frac{C}{r^2}\,,\ee where $K$ is  the mean extrinsic curvature and $C$ is an integration constant. For the foliation to approach the future event horizon and the future null infinity, we choose $K>0$ and $C>8m^3 K/3$.
The hyperboloidal CMC foliation is very well suited for our purposes and will be used
 throughout the rest of the paper (though many  expressions below are valid in any foliation~\eqref{time_traf}).
\subsection{Yang-Mills equations and the energy}
For the $SU(2)$ Yang-Mills potential $A$ we assume the spherically symmetric 'purely magnetic' ansatz \cite{d}
\begin{equation}\label{anstazA}
    A= W\, \tau_1\,d\vartheta+(\cot\vartheta\,\tau_3+W \,\tau_2) \sin\vartheta \,d\varphi\,,
\end{equation}
where $\tau_i$ are the generators of $SU(2)$  and $W=W(t,r)$. For this ansatz the Yang-Mills  curvature
$F=dA+[A,A]$ takes the form (hereafter $\dot{}=\partial_t$ and $'=\partial_r$)
\begin{equation}\label{F}
    F=\dot W \,dt\wedge\Omega+W' dr\wedge \Omega - (1-W^2)\, \tau_3\, d\vartheta \wedge \sin\vartheta\, d\phi\,,
\end{equation}
where $\Omega=\tau_1\, d\vartheta +\tau_2 \,\sin{\vartheta}\, d\varphi$, and
the Yang-Mills equation $\nabla_{\alpha} F^{\alpha\beta}+[A_{\alpha},F^{\alpha\beta}]=0$
on the Schwarzschild background with metric \eqref{eq:decomp} becomes a scalar semilinear wave equation
\be\label{ymeq}\frac{1-(NH)^2}{N} \ddot{W} + 2 N H \dot{W}' +(N H)'\dot W= \left(N W'\right)' + \frac{W(1-W^2)}{r^2}\,. \ee
We are interested in the Cauchy problem for smooth compactly supported
initial data. We stress that no boundary condition is imposed at the horizon  and the field $W(t,r_h)$ can evolve freely which, in particular, allows for dynamical connections between two vacuum states $W=\pm 1$. This behavior should be contrasted with the flat spacetime case where the vacuum states are disconnected since the value of field at the origin is rigidly fixed to $|W(t,r=0)|=1$ by the smoothness condition.

Note that Eq.\eqref{ymeq} has the reflection symmetry: if $W(t,r)$ is a solution, so is $-W(t,r)$. As we shall see below, this discrete symmetry is a key feature which shapes the structure of the phase space for Eq.\eqref{ymeq}.

Using the energy momentum tensor
\begin{equation}\label{enmom}
    T_{\alpha\beta}=\frac{1}{4\pi} \texttt{Tr}\left( F_{\alpha}^{\,\,\,\mu} F_{\mu}^{\,\,\,\beta}-\frac{1}{4} g_{\alpha\beta} F_{\mu\nu} F^{\mu\nu}\right)
\end{equation}
 and the timelike Killing vector field $\xi=\partial_t$, one can define the conserved current
\begin{equation}\label{J}
    J^{\alpha}=T_{\beta}^{\,\,\,\alpha}\, \xi^{\beta}\,,\qquad \nabla_{\alpha} J^{\alpha}=0\,.
\end{equation}
Integration of the current over a constant time spatial hypersurface $\Sigma_t$ yields the energy
\begin{equation}\label{en}
 E(t):=\int_{\Sigma_t} J^{\alpha} dS_{\alpha}=-4\pi \int_{r_h}^{\infty} T_t^{\,\,\,t}\, r^2 dr = \int_{r_h}^{\infty}  \left[ \frac{1-(NH)^2}{N}  \dot W^2 + N W'^2 + \frac{(1-W^2)^2}{2 r^2} \right] dr\, .
\end{equation}
Differentiating the energy with respect to time and using Eq.\eqref{ymeq} we get
\begin{equation}\label{gen_flux}
\dot{E} = 2 \left(-N H \dot{W}^2 + N W' \dot{W} \right)\Big|^{\infty}_{r_h}\,.
\end{equation}
In the hyperboloidal CMC coordinates \eqref{cmc} we have $H(r)\sim -N^{-1}$ near $r_h$ and $H(\infty)=1$, hence
\be \label{cmc_flux} \dot{E} = -2 \dot{W}^2(t,r_h) - 2 \dot{W}^2(t,\infty)\,. \ee
This expression shows that the energy is monotonically decreasing due to the energy flux through the horizon and through future null infinity.

As mentioned above, solutions starting from smooth initial data remain smooth for all future times and generically they scatter. However, the asymptotic completeness fails because there exist non-generic solutions which do not disperse. Our goal is to analyze the non-dispersive solutions in detail.  It follows from the monotonicity formula for the total energy \eqref{cmc_flux}  that the only possible mechanism of avoiding complete dispersion to vacuum is the stabilization of evolution on a nontrivial static solution. The discussion of static solutions is a subject of the next section.

We choose $m/2$ as a unit of length so hereafter $r_h=2m=1$.
\section{Static solutions and their perturbations}
\label{sec:static}
\subsection{Static solutions}
For static solutions Eq.\eqref{ymeq} reduces to the ordinary differential equation
\be\label{static} \left(1-\frac{1}{r}\right) W'' +\frac{1}{r^2} W' + \frac{W(1-W^2)}{r^2} = 0\,. \ee
Note that this is the Euler-Lagrange equation for the static energy functional
\begin{equation}\label{enst}
    E(W) = \int_{1}^{\infty}   \left[\left(1-\frac{1}{r}\right) W'^2 + \frac{(1-W^2)^2}{2 r^2} \right] dr\,.
\end{equation}
Solutions of Eq.\eqref{static} that are smooth at the horizon behave near $r=1$ as
\begin{equation}\label{ic}
    W(r)=a - a(1-a^2) (r-1) + \mathcal{O}\left((r-1)^2\right)\,,
\end{equation}
where $a$ is a free parameter.
It is not difficult to show using shooting methods that  there is a countable sequence of positive parameters $a_n$ ($n=0,1,...$) such that the corresponding solutions, denoted by $W_n(r)$, exist for all $r\geq 1$ and tend to $W_n(\infty)=(-1)^n$.  The  index $n$ denotes the number of nodes of the solution $W_n(r)$. These solutions can be obtained in the decoupling limit from colored black hole solutions of the coupled Einstein-Yang-Mills system \cite{b2} and the proof of their existence is implicit in the proof of existence of colored black holes given in \cite{bfm} (see also \cite{b1} for a variational argument). The solution $W_0=1$ is the ground state for which the energy has the global minimum $E(W_0)=0$. The solutions with indices $n>0$ can be viewed as excitations of the ground state; their energies $E_n=E(W_n)$ grow monotonically with $n$ ($E_1=0.47949, E_2=0.4994, E_3=0.4999,...$) and tend to the limit $E_{\infty}=1/2$ which is the energy of the singular solution $W_{\infty}=0$. Note that due to the reflection symmetry each static solution (except $W_{\infty}$) exists in two copies $\pm W_n$.
It is remarkable that the $n=1$ solution has been found in closed form \cite{bc}
\begin{equation}\label{w1}
    W_1(r)=\frac{c-r}{r+3(c-1)}\,,\qquad c=\frac{1}{2}(3+\sqrt{3})\,.
\end{equation}
\subsection{Linear perturbations}
\label{subsec:pert}
In order to understand the role of static solutions in the evolution we first need to examine their linearized stability. To this end
we substitute $W(t,r)=W_n(r)+w(t,r)$ into Eq.\eqref{ymeq}. Dropping quadratic and higher terms in $w$, we get the linear evolution equation for small perturbations about the static solution $W_n$
\be\label{pert}\frac{1-(NH)^2}{N} \ddot{w} + 2 N H \dot{w}' +(N H)'\dot w= \left(N w'\right)' + \frac{1-3W_n^2}{r^2}\,w\,, \ee
which after separation of variables, $w(t,r)=e^{\lambda t} v(r)$, leads to the eigenvalue problem
\be\label{eigen}\lambda^2 \frac{1-(NH)^2}{N}  v + 2 \lambda N H  v' + \lambda (N H)' v= \left(N v'\right)' + \frac{1-3W_n^2}{r^2}\,v\,. \ee
For the eigenvalues to come about we must identify the admissible behavior of eigenmodes at the endpoints $r=1$ and $r=\infty$. This rather subtle issue has a particularly simple answer in the hyperboloidal CMC foliation, where 'admissible' simply means 'smooth'.
The horizon $r=1$ is the regular singular point with the two linearly independent solutions
behaving as $v\sim (r-1)^{\alpha}$, where $\alpha=0$ or $-2\lambda$. Thus, assuming that $-2\lambda$ is not a positive integer (the absence of such algebraically special eigenvalues has been checked separately),  only the solution with $\alpha=0$ is admissible. Near the irregular singular point at infinity the admissible solution behaves as $v\sim 1$ (the second solution behaves as $v\sim e^{2\lambda r}$).

\noindent \emph{Remark.} In terms of the new dependent variable $u(r)=e^{-\lambda h(r)} v(r)$ (recall that $h(r)$ is the height function of the foliation \eqref{time_traf}) the eigenvalue problem \eqref{eigen} takes the standard Sturm-Liouville form (which is  the same as in Schwarzschild coordinates)
\begin{equation}\label{st}
 \mathcal{A} u =-\lambda^2 u\,,\qquad \mathcal{A}= -N \partial_r \left(N
    \partial_r\right)-\frac{N} {r^2} (1-3W_n^2)\,.
\end{equation}
The operator $\mathcal{A}$ is self-adjoint in the Hilbert space $X=L_2\left([1,\infty),N^{-1} dr \right)$. For $\text{Re}(\lambda)>0$ the admissible solution belongs to $X$, hence $\text{Re}(\lambda)>0$ implies that $\lambda^2$ is real, and therefore $\lambda$ is real as well.
However, for  $\text{Re}(\lambda)\leq 0$ the admissible solution  does not belong to $X$ so in this case no self-adjoint formulation is available and  eigenvalues are in general complex.
\vskip 0.2cm
Solving the eigenvalue problem (\ref{eigen}) numerically for the first few solutions $W_n$ we  found that the $n$th solution has exactly $n$ positive eigenvalues (hence $n$ unstable modes). Denoting the spectrum of eigenvalues for the solution $W_n$ by $\{\lambda_k^{(n)}\}$ and ordering it by a decreasing real part, we thus have
\begin{equation}\label{spectrum}
    \lambda_0^{(n)}> \lambda_1^{(n)}>\dots> \lambda_{n-1}^{(n)}>0> \text{Re}(\lambda_n^{(n)})>\text{Re}(\lambda_{n+1}^{(n)})>\dots
\end{equation}
Below we shall suppress the superscript $(n)$ on eigenvalues whenever it is clear from the context which static solution is considered.

  The  eigenfunctions with $\text{Re}(\lambda)<0$ are  called quasinormal modes.
  Numerical calculation of quasinormal modes is difficult because for $\text{Re}(\lambda)<0$ the "bad" solutions, $v\sim (r-1)^{-2\lambda}$ near $r=1$ and $v\sim e^{2\lambda r}$ near $r=\infty$, are subdominant with respect to smooth solutions and consequently it is hard to keep track thereof. For solutions that are known in closed form (like $W_0$ and $W_1$ in our case) an ingenious method of calculating quasinormal modes was developed by Leaver \cite{leaver}. In the case at hand this method works as follows. The mode which is smooth at the horizon is represented by a power series
  $v(z)=\sum a_n z^n$, where $z=(r-1)/r$. Since there are no singularities inside the circle $|z|<1$ in the complex plane, this power series expansion is absolutely convergent for $|z|<1$  but, in general, it diverges at  $z=1$. The quantization of eigenvalues comes from the condition of convergence of the power series  at $z=1$ which is fulfilled if and only if  the coefficients $a_n$ form a minimal solution of the recurrence relation. Due to Pincherle's theorem, relating the existence of a minimal solution to the convergence of an infinite continued fraction, this leads to a transcendental equation which is solved numerically. Using this method we calculated the first few eigenvalues for solutions $W_0$ and $W_1$. The results are shown in Table~\ref{table:qnm}.

\begin{table}[ht]
\begin{tabular}{|c|c|c|c|c|c|}
\hline $\,\,n\,\,$ & $\lambda_0^{(n)}$ & $\lambda_1^{(n)}$ & $\lambda_2^{(n)}$ & $\lambda_3^{(n)}$ & $\lambda_4^{(n)}$\\
\hline 0 & $-0.1849\pm 0.4965 i\,$ & $-0.5873\pm0.4290 i\,$ & $-1.0504\pm 0.3495 i\,$ & $-1.5438\pm 0.2923 i\,$ & $-2.0451\pm 0.2531 i$  \\
\hline 1 & $0.23243$ &$-0.0401 \pm 0.0422 i$  & $-0.6282 \pm 0.0139 i$  & $-1.0770   \pm 0.0214 i$ &$-1.5495 \pm 0.0223 i$\\
\hline\end{tabular}
  \caption{The first five eigenvalues  for linear perturbations about solutions $W_0$ and $W_1$. The eigenvalues $\lambda_k^{(0)}$  are well known in the literature as $\ell=1$ electromagnetic quasinormal modes \cite{leaver2}.
\label{table:qnm}}
\end{table}
\vspace{-1cm}
\section{Numerical results}
\label{sec:numeric}
\vspace{-0.3cm}
\subsection{Evolution of generic data}
We solve numerically the Cauchy problem for Eq.\eqref{ymeq} for a compactly supported gaussian pulse around the vacuum state $W=1$ with amplitude $p$. The numerical techniques are similar to those applied in \cite{z1}. We use the method of lines with a fourth order Runge-Kutta time integration and eighth order spatial finite differences. We employ hyperboloidal scri fixing coordinates  (\cite{z2}) based on the CMC foliation given in \eqref{cmc} with the parameters $K=0.5, C=0.5$. We use the radial coordinate $\rho=r/(r+1)$, which compactifies  the exterior Schwarzschild region $1\leq r<\infty$ into the finite interval $1/2\leq \rho\leq 1$.  There are no incoming characteristics into the simulation domain, therefore no boundary conditions are applied. At the boundaries of the simulation domain we use one-sided finite differencing. We refer an interested reader to \cite{z1} for more details about the numerical method and the concrete form of the symmetric hyperbolic system corresponding to Eq.\eqref{ymeq}.

For generic amplitudes the energy in any bounded region decays to zero and the solution approaches one of the vacuum states $W=\pm 1$. The quantitative description of the late stages of this relaxation process has been given in \cite{bcr,z1}. For intermediate times the decay to vacuum has the form of quasinormal ringing  with the fundamental eigenvalue $\lambda_0=-0.1849+0.4965 i$. For late times, the nonlinear tail decaying as $t^{-4}$ becomes uncovered.

\begin{figure}[ht]
\center
\includegraphics[width=0.44\textwidth]{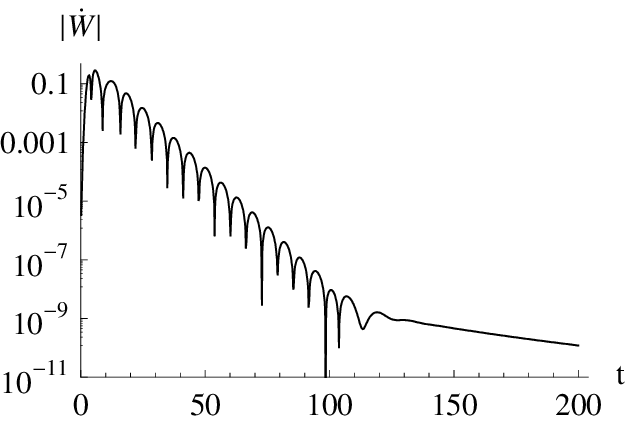}\hspace{1cm}
\includegraphics[width=0.44\textwidth]{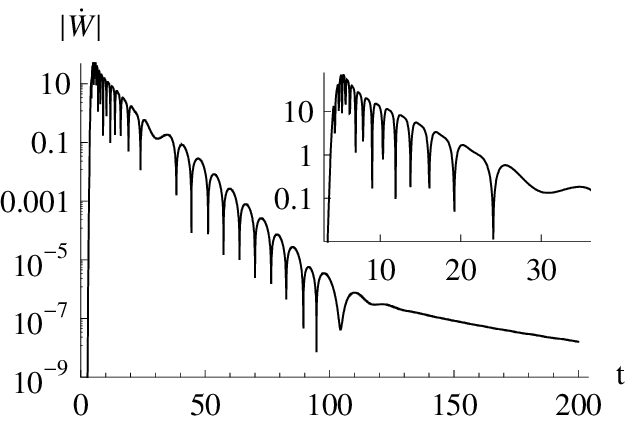}
\caption{The evolution of generic initial data (observed at $\rho_\mathrm{obs}=2/3$) with small (left) and large (right) energy. After an initial transient phase depending on initial data, the solution goes through a phase of exponentially damped oscillations (quasinormal ringing) and then a polynomial decay (tail) to one of the vacuum states. The first quasinormal mode  has the eigenvalue $\lambda_0=-0.1849+0.4965 i$. The tail falls off as $t^{-4}$. For large energy the initial transient phase takes longer: after the direct signal from the data passes through, there appear nonlinear oscillations with exponentially decreasing amplitude and frequency (see the inset on the right plot).
\label{fig:generic}}
\end{figure}
 Let us point out in passing an interesting difference between the evolution of small and large energy initial data, as shown in Fig.~\ref{fig:generic}. Namely,
for large energy  data, after the direct signal passes through but before the ringdown, there is a clearly pronounced  phase of evolution during which an excess energy is radiated away in the form of nonlinear oscillations with exponentially decreasing amplitude and frequency (see the inset on the right plot in Fig.~\ref{fig:generic}). The time span of this phase increases with energy. Such nonlinear oscillations are characteristic for large energy solutions of defocusing nonlinear wave equations and in our opinion they deserve further investigation, but it will not be pursued here.

\begin{figure}[ht]
\center
\includegraphics[height=0.23\textheight,width=0.48\textwidth]{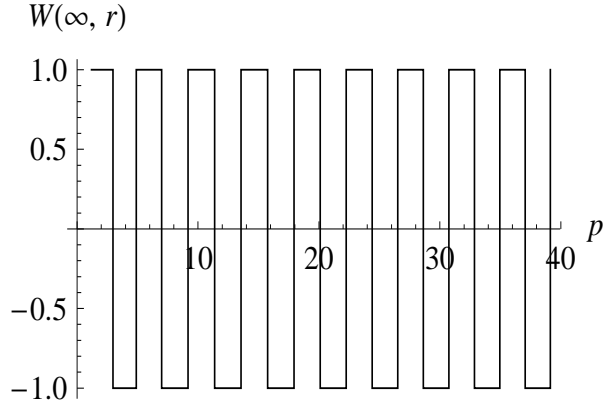}
\caption{The vacuum endstate of evolution as a function of initial amplitude $p$. The plot indicates that the flip of vacuum state is asymptotically periodic in $p$ (with the period  depending on a family of initial data).
\label{fig:distr}}
\end{figure}

 Since initial datum is a  perturbation of $W=1$, for small amplitudes $p$ all solutions tend asymptotically to $W=1$. However, as $p$ grows, the endstate flips back and forth from one vacuum state to another, which indicates that the curve of initial data repeatedly intersects a borderline between basins of attractions of two vacuum states.
Fig.~\ref{fig:distr} shows the final state of evolution against the amplitude of initial data. The parameter space is partitioned into windows of generic evolution. Somewhat surprisingly, the appearance of these windows seems asymptotically periodic in $p$ (although the period is not universal).

A natural question is what happens at the borderline between basins of attraction of two vacuum states. This question is the subject of the next section.

\subsection{Approach to and departure from the static solution $W_1$}
In this section we give a \emph{quantitative} description of non-generic solutions which do not disperse. Using bisection we fine-tune the amplitude to one of the critical values of the amplitude, which we denote by $p^*$. The evolution of such nearly critical initial data exhibits a universal intermediate phase during which the solution hangs around  the static solution $W_1$, first approaching it and then departing from it. This is a typical behavior around a saddle point. In other words, the solution $W_1$ plays the role of an intermediate attractor and its codimension-one stable manifold separates the basins of attraction of two vacuum states.

\begin{figure}[ht]
\center
\includegraphics[width=0.46\textwidth]{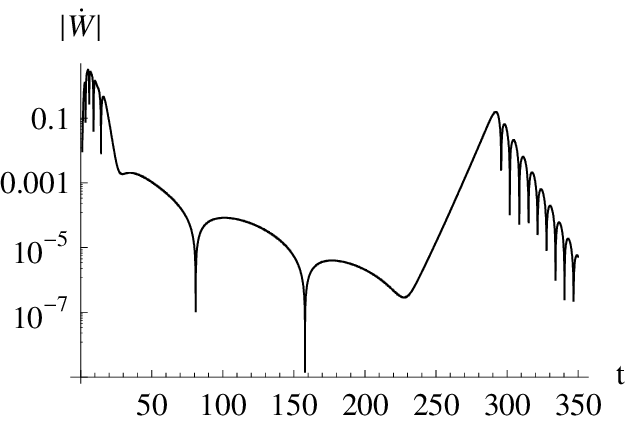}\hspace{0.8cm}
\includegraphics[width=0.46\textwidth]{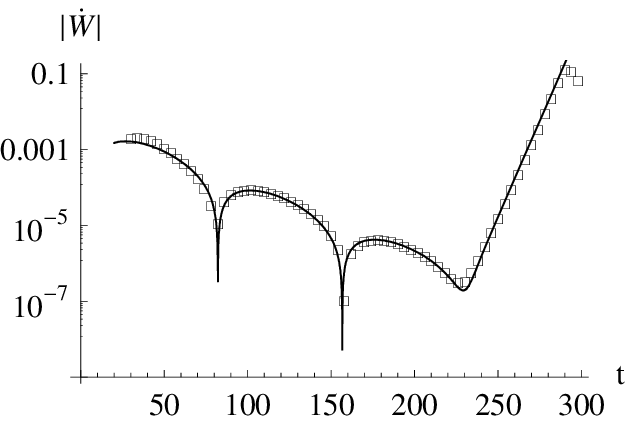}
\caption{The evolution of initial data fine-tuned in quadruple precision to the border between basins of attraction of two different vacuum states. To show pointwise convergence to and departure from the static solution $W_1$ we plot the time derivative of the Yang-Mills potential $\dot W(t,\rho_0)$ at an arbitrarily chosen $\rho_\mathrm{obs}=2/3$ in half-logarithmic scale. On the left panel, after an initial transient signal coming from initial data, one can distinguish three universal phases of evolution: quasinormal ringdown to the attractor, exponential departure from the attractor, and finally an approach to one of the vacuum states.  On the right panel we fit the function \eqref{inter} to the numerical solution (small squares). Only the parameters $c$, $A$, and $\delta$ are fitted; the remaining parameters $\text{Re}(\lambda_1)=-0.040103$, $\text{Im}(\lambda_1) = 0.042173$, and $\lambda_0=0.23243$
are supplied  by the linear stability analysis from Section~\ref{subsec:pert}.
\label{fig:inter}}
\end{figure}

The evolution near the intermediate attractor can be described quantitatively using the results from Section~\ref{subsec:pert}. We claim that for intermediate times the following approximation is valid
\begin{equation}\label{inter}
    W(t,\rho)-W_1(\rho) \simeq c\, (p-p^*)\, e^{\lambda_0 t}\, v_0(\rho)+ A \,e^{-|\text{Re}(\lambda_1)| t} \sin(\text{Im}(\lambda_1) t+\delta)\, v_1(\rho)\,,
\end{equation}
where $v_0(\rho)$ and $v_1(\rho)$ are, respectively, the eigenfunction of the single unstable mode with eigenvalue $\lambda_0=0.23243$ and the first quasinormal mode with the eigenvalue $\lambda_1=-0.040103+0.042173 i$. Other quasinormal modes  are not included in \eqref{inter} because they are damped much faster (see Table~\ref{table:qnm}).
In principle, the expression \eqref{inter} should also include the contribution from the  tail, however, for the intermediate times involved in our simulations, the tail is negligible in comparison with the quasinormal mode so we omit it. The numerical verification of the approximation \eqref{inter} is shown in Fig.~\ref{fig:inter} where we plot the result of a quadruple precision bisection study.

\begin{figure}[ht]
\center
\includegraphics[width=0.48\textwidth]{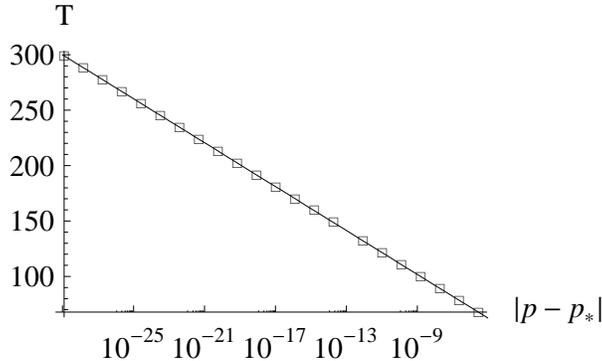}
\caption{The lifetime $T$ of the intermediate attractor $W_1$ as a function of the distance from the critical amplitude in a half-logarithmic plot. The small squares correspond to numerical evolutions. The simple least square fit depicted by the thin line gives $\lambda_\mathrm{fit}= 0.23246$, in excellent agreement with the formula \eqref{life} and the eigenvalue $\lambda_0 = 0.23243$ obtained via linear stability analysis.
\label{fig:lifetime}}
\end{figure}

We can define a "lifetime" $T$ of the intermediate attractor  as a span of time during which the solution stays in some given neighborhood of the static solution $W_1$. The lifetime is determined by the time in which the amplitude of the unstable mode grows to a given size, that is $|p-p^*| e^{\lambda_0 T} =O(1)$, which yields
\begin{equation}\label{life}
    T\sim -\frac{1}{\lambda_0} \ln|p-p^*|\,\quad \mbox{as}\quad  p\rightarrow p^*\,.
\end{equation}
The numerical verification of this scaling law is shown in Fig.~\ref{fig:lifetime}.

It follows from \eqref{life}  that  during the lifetime $T$ the first quasinormal mode decays by the factor
 \begin{equation}\label{dist}
e^{-|\text{Re}(\lambda_1)| T} \sim |p-p^*|^{\frac{|\text{Re}(\lambda_1)|}{\lambda_0}}\,,
\end{equation}
hence, for a given precision of bisection, the closest approach to the unstable attractor is determined by the ratio of the damping rate of the first quasinormal mode (which governs the rate of convergence to the attractor) and the eigenvalue of the unstable mode (which governs the rate of departure from the attractor).

\subsection{Nonlinear bi-instability of $W_1$}

\begin{figure}[ht]
\center
\includegraphics[width=0.44\textwidth]{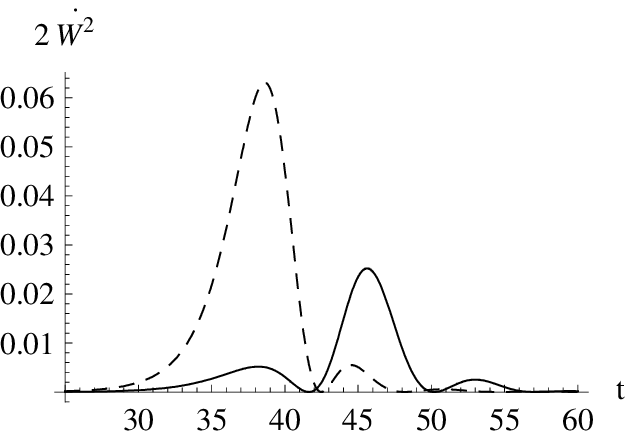}\hspace{1cm}
\includegraphics[width=0.44\textwidth]{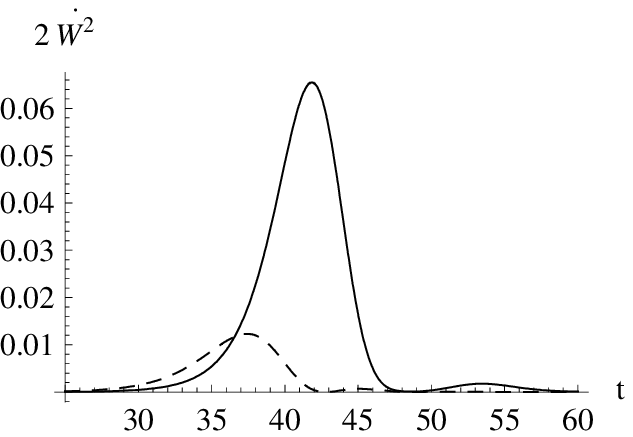}
\caption{Energy fluxes  through the event horizon (dashed curve) and through scri (solid curve) during the nonlinear decay of the static solution $W_1$. The endstate of evolution is $W=-1$ (left) and $W=+1$ (right). The corresponding integrated energy fluxes are given in Table \ref{table:energy}.
\label{fig:flux}}
\end{figure}

Having described the evolution for intermediate times we turn now to the description of  nonlinear decay of the intermediate attractor to one of the vacuum states. We ask: what is the ratio of energy that falls into the black hole to the energy that disperses to infinity? To study this question accurately we do not perform bisection, but start the evolution from initial data having the form of the solution $W_1$ plus a small perturbation.
Depending on the direction of the perturbation, the endstate of evolution is $W=-1$ or $W=+1$. Note that in the hyperboloidal scri fixing foliation (in contrast to foliations whose leaves meet at spatial infinity), the asymptotic value of the filed at scri, $W(t,\rho=1)$, can change continuously during the evolution, in particular the vacuum state at scri can change from one to another.
In the case at hand we have $W(0,\rho=1)=-1$ (since $W_1(\rho=1)=-1$ and the small perturbation is compactly supported).
Using the expression \eqref{cmc_flux} we compute the energy flux through the horizon and through scri during the nonlinear decay of $W_1$ until the time $t=200$. The  results of this computation are shown in Fig.~\ref{fig:flux} and Table~\ref{table:energy}.

\begin{table}[ht]
\begin{tabular}{|c|c|c|c|c|}
\hline Endstate & $E_{\mathrm{horizon}}$ &$E_{\mathrm{scri}}$ & $E_{\mathrm{radiated}} $ &  $ E_\mathrm{initial} - E_{\mathrm{radiated}} $ \\
\hline $-1$ & $\,0.3422522541\,$  & $\,0.1372383912\,$  & $\,0.479490645\,$  &  $+ 1.9 \times 10^{-8}$ \\
\hline $+1$ & $\,0.0851824567\,$  & $\,0.3943078254\,$  & $\,0.479490282\,$  & $+ 3.8 \times 10^{-7} $\\
\hline\end{tabular}
  \caption{The amount of energy radiated through the horizon and through scri obtained by integration of energy fluxes shown in Fig.~5 up to $t=200$. The initial energy of the static solution $W_1$ plus a small perturbation is $E_\mathrm{initial}= 0.479490664$.
\label{table:energy}}
\end{table}

We find that if the endstate is $W=-1$ (no change of the vacuum state at scri) then most of the total energy falls into the black hole ($71\%$), while if the endstate is $W=+1$ most of the energy escapes to infinity ($82\%$).
 The total balance of radiated energy calculated numerically is accurate up to $10^{-7}$, which is very reassuring. The remaining tiny fraction of energy is radiated for $t>200$ in the form of a tail. Since the tail at the horizon falls off faster than the tail along scri,
 the error shown in the last column of Table~\ref{table:energy} is larger for the endstate $+1$ where
 most of the energy is radiated to infinity.

\section{Discussion}
\label{sec:discuss}
 The dynamics of the Yang-Mills field on four dimensional Minkowski spacetime is rather indistinctive as all solutions evolve in the same manner dispersing asymptotically to vacuum.
 We hope to have convinced the reader that in the case of Schwarzschild background the dynamics is much more interesting. This  is due to the presence of the horizon which affects the Cauchy problem in several respects.
 First, the horizon breaks scale invariance and thereby allows for existence of nontrivial static solutions. Second, the horizon makes the phase space simply connected, in particular the two vacuum states $W=\pm 1$ are homotopic which makes it possible to perform bisection between their basins of attraction (in contrast to flat space where the two vacuum states are separated by an infinite energy barrier). Finally, the horizon acts as an absorption boundary and thus provides an additional (besides dispersion to infinity) mechanism of dissipation of energy.

  All the above features, combined with the fact that the static solution with single unstable mode is known is closed form, make  the Yang-Mills equation on the Schwarzschild background an attractive toy-model for gaining better understanding of codimension-one stationary attractors for nonlinear wave equations.
  Previous studies of such attractors, performed mainly in the context of Type I critical phenomena in gravitational collapse (see, e.g. \cite{ccb,bc2}), have focused on the dynamics of departure from the attractor along the unstable manifold\footnote{A notable exception is Master's thesis of N. Szpak \cite{nik} in which he analyzed the saddle-point dynamics around the unstable static solution for the focusing semilinear wave equation $u_{tt}-\Delta u -u^5=0$ in four dimensional Minkowski spacetime and conjectured that the rate of convergence is exponential for intermediate times, however as yet the alleged purely damped quasinormal mode has not been confirmed in a perturbative calculation.}. To our knowledge, the present paper is the first one where convergence
  to an unstable stationary attractor has been shown to proceed via quasinormal ringing
  for intermediate times. Apart from the theoretical importance, this result has practical implications for numerical searches of unstable attractors, as follows from the formula~\eqref{dist}.
\vskip 0.2cm \noindent \textbf{Acknowledgments:} This research was supported in part by the MNII grants: NN202 079235 and 189/6.PRUE/2007/7 and by the Marie Curie Transfer of Knowledge contract MTKD-CT-2006-042360.

\end{document}